\documentclass{aa}  

\usepackage{amsmath}	% Advanced maths commands
\usepackage{amssymb}	% Extra maths symbols
\usepackage{gensymb}
\usepackage{enumitem}
\usepackage{subcaption}
\usepackage{natbib,twoopt}
\usepackage{array}
\usepackage{graphicx}
\usepackage[colorlinks,citecolor=blue,linkcolor=blue,urlcolor=blue,bookmarks=false,hypertexnames=true]{hyperref}

\usepackage{txfonts}
\usepackage[T1]{fontenc}
\usepackage{ae,aecompl}

\newcommand{\swift}{{\it Swift}}

\begin{document}

\title{Exploring the accretion-ejection geometry of GRS 1915+105 in the obscured state with future X-ray spectro-polarimetry}
   \author{A. Ratheesh\inst{1,2},
          G. Matt\inst{3},
          F. Tombesi\inst{1,4,5,6,7},
          P. Soffitta\inst{2},
          M. Pesce-Rollins\inst{8}, and 
          A. Di Marco\inst{2}}
          %X. YY\inst{8}}
          
    \institute{Department of Physics, Tor Vergata University of Rome, Via della Ricerca Scientifica 1, I-00133 Rome, Italy\\
              \email{ajay.ratheesh@roma2.infn.it}
         \and
             INAF - IAPS, Via Fosso del Cavaliere 100, I-00133 Rome, Italy
        \and Dipartimento di Matematica e Fisica, Universita degli Studi Roma Tre, Via della Vasca Navale 84, I-00146 Roma, Italy
        \and INAF - Astronomical Observatory of Rome, Via Frascati 33, I-00078 Monte Porzio Catone (Rome), Italy 
        \and INFN - Tor Vergata, Via della Ricerca Scientifica 1, I-00133 Rome, Italy
        \and Department of Astronomy, University of Maryland, College Park, MD 20742, USA
        \and NASA/Goddard Space Flight Center, Greenbelt, MD 20771, USA
        \and INFN-Pisa, Largo B. Pontecorvo 3, 56127 Pisa, Italy
        }

\date{}

% \abstract{}{}{}{}{} 
% 5 {} token are mandatory
 
  \abstract
   {GRS 1915+105 has been in a bright flux state for more than 2 decades, but in 2018 a significant drop in flux was observed, partly due to changes in the central engine along with increased X-ray absorption.}
   {The aim of this work is to explore how X-ray spectro-polarimetry can be used to derive the basic geometrical properties of the absorbing and reflecting matter. In particular, the expected polarisation of the radiation reflected off the disc and the putative outflow is calculated.}
   {We use \textit{NuSTAR} data collected after the flux drop to derive the parameters of the system from hard X-ray spectroscopy. The spectroscopic parameters are then used to derive the expected polarimetric signal, using results from a MonteCarlo radiative transfer code both in the case of neutral and fully ionised matter.}
   {From the spectral analysis, we find that the continuum emission becomes softer with increasing flux, and that in all flux levels the obscuring matter is highly ionised. This analysis, on the other hand,  confirms that spectroscopy alone is unable to put constraints on the geometry of the reflectors. Simulations show that X-ray polarimetric observations, like those that will be provided soon by the Imaging X-ray Polarimetry Explorer (IXPE), will help  determining the geometrical parameters which are left unconstrained by the spectroscopic analysis.}
   {}
   
\keywords{Accretion, accretion discs -- stars: winds, outflows -- X-rays: binaries, -- Polarization, -- Relativistic processes }

\titlerunning{Spectro-Polarimetry of obscured state of GRS 1915+105}
\authorrunning{A, Ratheesh et al}
\maketitle
\section{Introduction}
GRS 1915+105 is a low mass X-ray Binary (LMXB), hosting a black hole of mass 12.4 M$_\odot$ and a K1 star \citep{Boeer_1996A&A...305..835B,Reid_2014ApJ...796....2R}. The high luminosity and the extreme variability in the light curves and spectra seen \citep{Belloni_2000A&A...355..271B} in this system make it a unique source to study the accretion-ejection phenomena in galactic black hole binaries (BHBs). \\

State changes are a common feature in BHBs. In general the spectral states can be classified as high/soft (HS) and low/hard (LH) \citep{Belloni_2000A&A...355..271B}. Some BHBs, especially transient ones, exhibit hysteresis in their hardness intensity diagram (HID), initially rising from LH state and later transiting towards HS state \citep{Fender_2004}. At times, a jet is observed in the radio band during the hard state and also during the transition to soft state \citep{Fender_2004}. GRS 1915+105 properties differ from that of other transient BHBs, probably due to a large accretion disk. Even though GRS 1915+105 does not exhibit hysteresis in HID, like persistent BHBs such as Cygnus X-1, Cygnus X-3, hard-soft transitions are seen \citep{Belloni_2000A&A...355..271B,Vrtilek_2013MNRAS.428.3693V,Zdziarski_2016MNRAS.456..775Z}. An anti correlation between jet and accretion disc wind was reported in GRS 1915+105 \citep{2009Natur.458..481N}, such that in observations where wind was present the jet was weak or absent and in states where jet was present the wind was weak or absent. This suggested that the disc outflows in the form of winds could play a role in the state transitions in these systems. The state dependence of the accretion disc wind is not clearly understood at present, as there exist multiple mechanisms responsible for a disc wind and the predominant mechanism might differ depending on the state. In GRS 1915+105, magneto hydro-dynamical modelling (MHD) of the wind indicated that the density of the wind in a soft state is two orders of magnitude higher than in a hard state \citep{Ratheesh_2021}. A similar analysis in some other BHBs are currently under progress \citep{Fukumura_2020AAS...23620801F}. Hence, to probe the role played by outflows in state transitions needs a clear understanding of the disc-jet-wind connection in different states. \\ 

Interestingly, since March 2018, GRS 1915+105 entered into a novel state. In fact, the source intensity started to decline gradually, while the hardness ratio increased indicating a spectral hardening \citep{Koljonen_2020A&A...639A..13K}. The source showed presence of absorption lines during this period \citep{Miller_2020arXiv200707005M, Balakrishnan_2020arXiv201215033B}. Further in 2019, the flux dropped sharply into a obscured state (OS) with reflected emission lines seen in the X-ray spectra \citep{Miller_2020arXiv200707005M}. In this faint state, a clear increase in the intrinsic absorption is apparent, making at times the source a Galactic equivalent of highly absorbed Active Galactic Nuclei \citep{Miller_2020arXiv200707005M}. Heavy absorption seen in high mass X-ray binaries (HMXBs) is understood to be from the stellar wind \citep{Kaper_1993A&A...279..485K,Oskinova_2012MNRAS.421.2820O}. However, in the case of low mass X-ray binaries (LMXBs), the absorption could be originating from the environment around the compact object making this a peculiar state in the accretion-ejection regime. 
There are different sources of obscuration in X-ray binaries, such as slim disks \citep{Motta_2017MNRAS.471.1797M}, optically thick accretion wind \citep{Middleton_2018arXiv181010518M, Miller_2020arXiv200707005M}, vertically extended outer disk \citep[GRS 1915+105: ][]{Neilsen_2020ApJ...902..152N}, etc. One of the main possible candidate in GRS 1915+105 could be a failed magnetic wind \citep{Miller_2020arXiv200707005M}, as magnetic winds are seen close to the central engine in X-ray binaries \citep{Miller2015_2015ApJ...814...87M,Miller2016_2016ApJ...821L...9M,2017NatAs...1E..62F, Ratheesh_2021}. \\

GRS 1915+105 has a high inclination angle of 66$^{\circ}$ $\pm$ 2$^{\circ}$ \citep{Fender_1999MNRAS.304..865F} with a large accretion disc, making it an ideal system to study the reprocessing of the continuum in the accretion disc winds. Accretion disc winds are preferentially seen at high inclination angles, indicating high column density in the equatorial direction \citep{Ponti_2012}. GRS 1915+105 is known to have very strong absorption lines in its X-ray spectra \citep{Ueda2009_2009ApJ...695..888U,2009Natur.458..481N,Miller2015_2015ApJ...814...87M,Miller2016_2016ApJ...821L...9M}. While transiting into the OS, the source was also seen in an intermediate state showing the presence of strong absorption lines predominantly at 6 to 7 keV \citep{Koljonen_2020A&A...639A..13K,Miller_2020arXiv200707005M,Neilsen_2020ApJ...902..152N}. The obscured state (OS) has strong emission lines \citep{Miller_2020arXiv200707005M}, indicating reflection from either a distant disc, wind or a torus. Spectroscopic data suggest that the obscurer could be a failed disc wind \citep{Miller_2020arXiv200707005M}. In OS, GRS 1915+105 has a flux of an order of magnitude lower than in the previous states. Sporadic X-ray flares are also observed in this state \citep{TOO_Iwakiri_2019ATel12787....1I,TOO_Neilsen_2019ATel12793....1N,TOO_Jithesh_2019ATel12805....1J}, unlike the kind of variability seen in the source before. Strong radio flares are also seen during this unusual state, even though any correlation of X-ray and radio flares are currently under investigation \citep{TOO_Motta_2019ATel12773....1M,TOO_Trushkin_2020ATel13442....1T}. \cite{Koljonen_2020A&A...639A..13K} indicates that the source becomes partially obscured in this state with a hardening of the spectral index, as also suggested by \cite{Miller_2020arXiv200707005M} indicating a change in the central engine properties. \cite{Koljonen_2020A&A...639A..13K} further suggest that the source could be going into a hard state with a low Eddington luminosity of 1-2 $\%$ but the strong radio flares suggest that the accretion rate has not reduced to a large extent. However, high radiative efficiency could be another possibility for the radio flares instead of the change in accretion rate. \\

It has long been recognised that the structure of the absorbing matter in AGN, and in particular in Compton-Thick ones, can be constrained by X-ray polarimetry \citep[e.g.][]{Goosmann_Matt_2011MNRAS.415.3119G}. With the upcoming launch of NASA/ASI's Imaging X-ray Polarimetry Explorer (IXPE) in late 2021, the X-ray polarimetric window will be opened \citep{Weisskopf_2016SPIE.9905E..17W}. The gas pixel detectors on-board IXPE will be sensitive to polarisation in the 2 to 8 keV range \citep{Costa_2001Natur.411..662C}. Polarimetry promises to break model degeneracies remaining after spectral and timing models analyses. In fact, different physical and geometrical models can produce similar spectral shapes, however their signatures in polarisation might be different. For example a spherical or a slab geometry of the corona of a compact object would have significantly different polarisation properties
\citep{Schnittman_2010ApJ...712..908S}. Similarly the geometry of the reflecting matter can also be probed using polarimetry. \\

Estimating the continuum emission in the OS requires broadband hard X-ray spectroscopy up to a few tens of keV as provided by \textit{NuSTAR}, which will also help in disentangling the continuum from the reflection features. In this paper we analyse \textit{NuSTAR} observations of GRS1915+105 taken after the source entered into the OS, to derive the main physical properties of the obscuring matter, as well as its geometrical parameters as far as can be constrained by spectroscopy alone.
We perform a flux resolved spectral analysis of the OS of GRS 1915+105 using relativistic reflection and absorption models. We also derive the expected polarimetric properties in X-rays for various spectral and geometrical parameters and then apply it to the case of the OS of GRS 1915+105. This paper is organised in the following manner. In section 2 we outline the justification for the selection of \textit{NuSTAR} data and data reduction methods. Section 3 contains results of the spectral analysis of \textit{NuSTAR} data initially for a non-flaring time intervals and then at different flux levels. Section 4 contains the results of polarimetric simulations using radiative transfer code for a toy model and derive its applications to the OS of this source. In section 5, we summarise and discuss the findings from this work, compare our results with previous ones, and present and discuss the feasibility of IXPE observations to constrain the geometry of the system.\\

\section{Data selection and reduction}
\begin{figure}[b]
\centering
\resizebox{\hsize}{!}{\includegraphics[width=1.2\hsize]{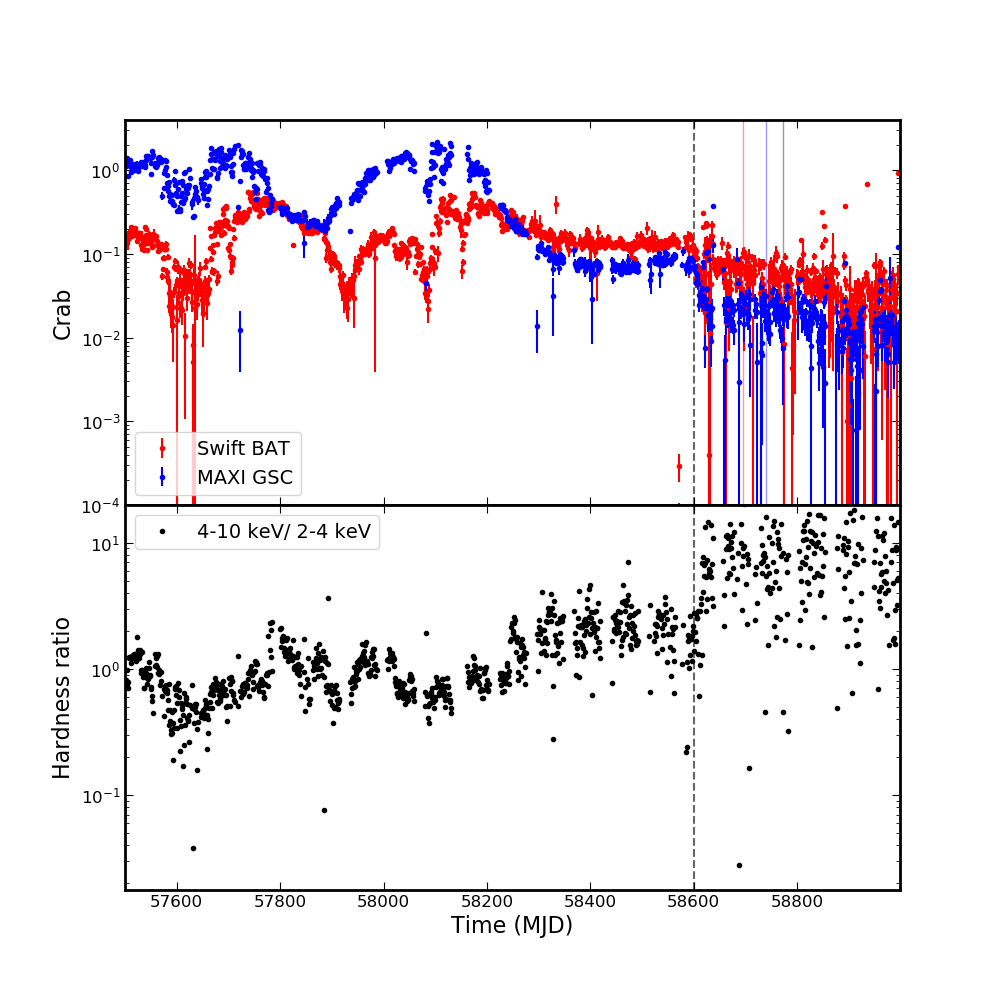}}
\caption{Top: \textit{\swift/BAT} and \textit{MAXI/GSC} daily averaged lightcurves in energy range 15-50 keV and 2-20 keV respectively. The red, blue and black vertical lines denote the \textit{NuSTAR} observations used in this work. The black dotted line indicates roughly the epoch at which the source enters the obscured state (OS). The flux is normalised to crab units. Bottom: The evolution of hardness ratio (4-10 keV/2-4 keV) with time in \textit{MAXI/GSC}.}
\label{MAXI_BAT_lc_hardness}
\end{figure}
\begin{figure}[b]
\resizebox{\hsize}{!}{\includegraphics[width=1.2\hsize]{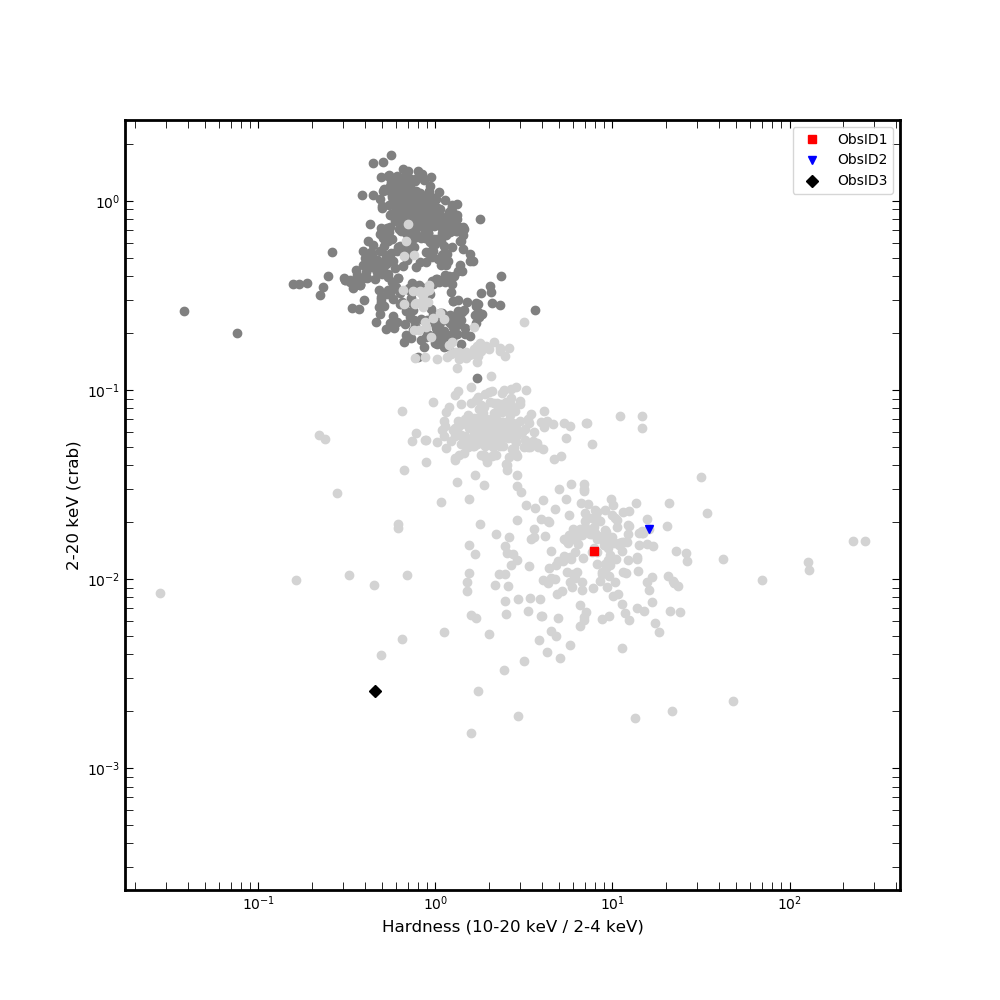}}
\caption{The hardness intensity diagram (HID) made using daily averaged \textit{MAXI/GSC} lightcurves. The black and grey points indicate the data points before and after transition epoch. The coloured points are during the MJD of the observations used in this work.}
\label{MAXI_HID}
\end{figure}
Daily monitoring of the X-ray sky has been possible due to the Burst Alert Telescope on board Neil Gehrels Swift Observatory \citep[\textit{\swift/BAT}:][]{SwiftBAT_2013ApJS..209...14K} and Gas Slit Camera on board Monitor of All-sky X-ray Image \citep[\textit{MAXI/GSC}:][]{Maxi_2009PASJ...61..999M} in hard (15-50 keV) and soft X-ray (2-20 keV) bands respectively. We used the \textit{\swift/BAT} lightcurves (15-50 keV) obtained from the \textit{\swift/BAT} transient database \footnote{\url{https://swift.gsfc.nasa.gov/results/transients/}} and \textit{MAXI/GSC} lightcurves (2-20 keV) obtained from \textit{MAXI/GSC} web-page \footnote{\url{http://maxi.riken.jp}} of GRS 1915+105 to determine the epoch at which the source faded into a low flux state (Fig. \ref{MAXI_BAT_lc_hardness}). The lightcurves plotted in Fig \ref{MAXI_BAT_lc_hardness} are normalised by the Crab count-rate in the respective energy bands. The Crab count-rates for energies 2-20 keV, 2-4 keV and 10-20 keV in \textit{MAXI/GSC} are 3.43 c s$^{-1}$ cm$^{-2}$, 1.90 c s$^{-1}$ cm$^{-2}$ and 0.36 c s$^{-1}$ cm$^{-2}$ respectively, and for energy 15-50 keV in \textit{\swift/BAT} is 0.22 c s$^{-1}$ cm$^{-2}$. The flux dropped around MJD 58600 and the change in the flux is also linked with the hardening of the spectra, which is seen in the hardness intensity diagram (HID) in Fig. \ref{MAXI_HID}. In OS, the source count-rate in the \textit{MAXI/GSC} is 0.01 to 0.03 Crab with hardness ratio greater than 2.5. During the transition to a heavily obscured state, the source was also in a partially obscured state with the presence of a strong absorption line as previously reported by \cite{Koljonen_2020A&A...639A..13K,Neilsen_2020ApJ...902..152N}. \\

\begin{figure}[ht!]
\centering
\resizebox{\hsize}{!}{\includegraphics[width=1.5\hsize]{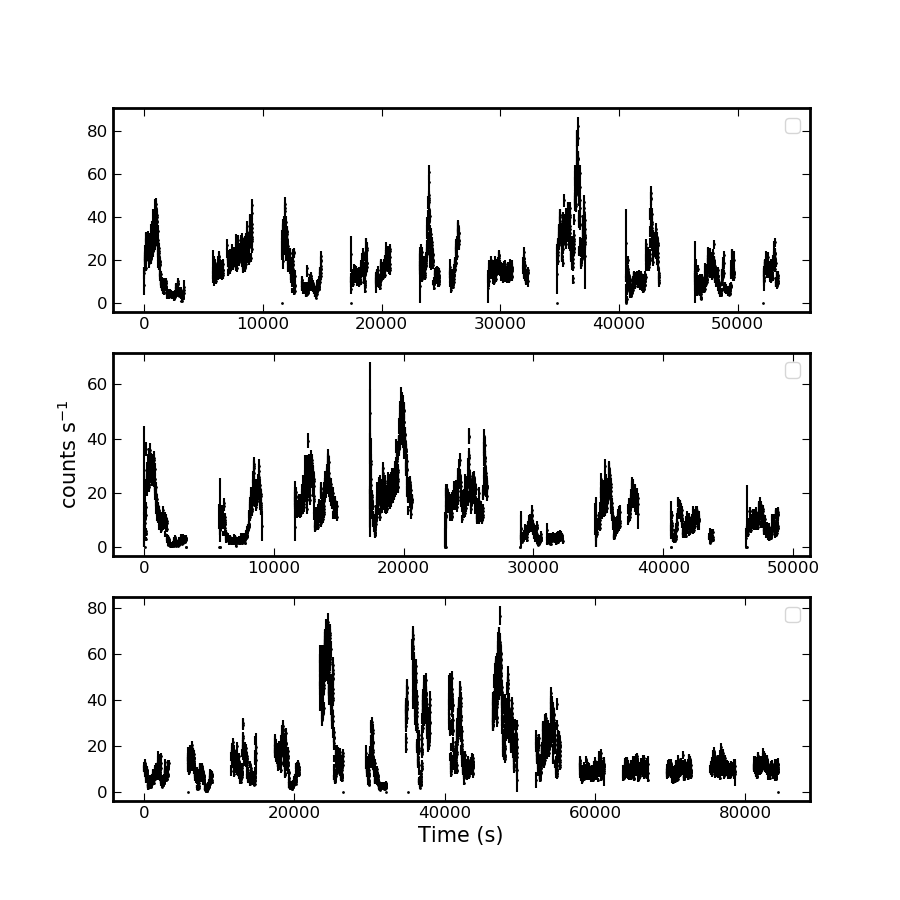}}
\caption{The \textit{NuSTAR/FPMA} lightcurves in 3$-$79 keV energy band. Top, middle and bottom represent obsID1, obsID2, and obsID3.  The lightcurve is plotted from start time of the observation and is binned at 10s.}
\label{nustar_lc}
\end{figure}

\textit{NuSTAR} observed GRS 1915+105 multiple times after the state transition into a low flux state. However in this work we concentrate on only three archived target of opportunity (ToO) observations taken while the source was in extremely absorbed state. The observations used in this analysis were performed on 2019-07-31 (\textit{NuSTAR} obsID:30502008004, from now on obsID1), 2019-09-13 (\textit{NuSTAR} obsID:30502008006, from now on obsID2) and 2019-10-16 (\textit{NuSTAR} obsID:90501346002, from now on obsID3) for an exposure of 23.2 ks, 23.7 ks, and 42.6 ks respectively. The red, blue and green vertical lines on the top panel of Fig. \ref{MAXI_BAT_lc_hardness} represents the MJD of obsID1, obsID2, obsID3. They are also marked as distinct points in Fig. \ref{MAXI_HID} from the \textit{MAXI/GSC} data on the MJD of these observations. The \textit{NuSTAR} data of GRS 1915+105 are reduced using v.0.4.6 of the NuSTARDAS pipeline with \textit{NuSTAR} CALDB v.20200526. The \textit{nupipeline} tool was used to generate cleaned event files. The source and the background for further analysis was selected manually using a circular region in the image using DS9 \citep{DS9_2003ASPC..295..489J}. The source region was centred at source location, and the background was selected away from the source to avoid contamination from the source. The radius of the circular region was determined manually to include maximum source photons in case of the source and to exclude source contamination in case of background (radius used for obsID1 is 57$^{''}$, obsID2 is 55$^{''}$ and obsID3 is 56$^{''}$). The \textit{nuproducts} tool was used to extract source background subtracted spectrum and lightcurve for the selected regions. The lightcurves of these observations from both FPMA and FPMB in 3$-$79 keV energy range and binned at 10 seconds are shown in Fig. \ref{nustar_lc}. The lightcurves show large variation in count rate as also previously reported by \cite{TOO_Iwakiri_2019ATel12787....1I,TOO_Neilsen_2019ATel12793....1N,TOO_Jithesh_2019ATel12805....1J}.\\
%The random flares seen in these lightcuves are different from the kind of variability observed in the source before \textbf{\citep{Belloni_2000A&A...355..271B}}. \\

\begin{figure}[ht!]
\centering
\resizebox{\hsize}{!}{\includegraphics[width=1.5\hsize]{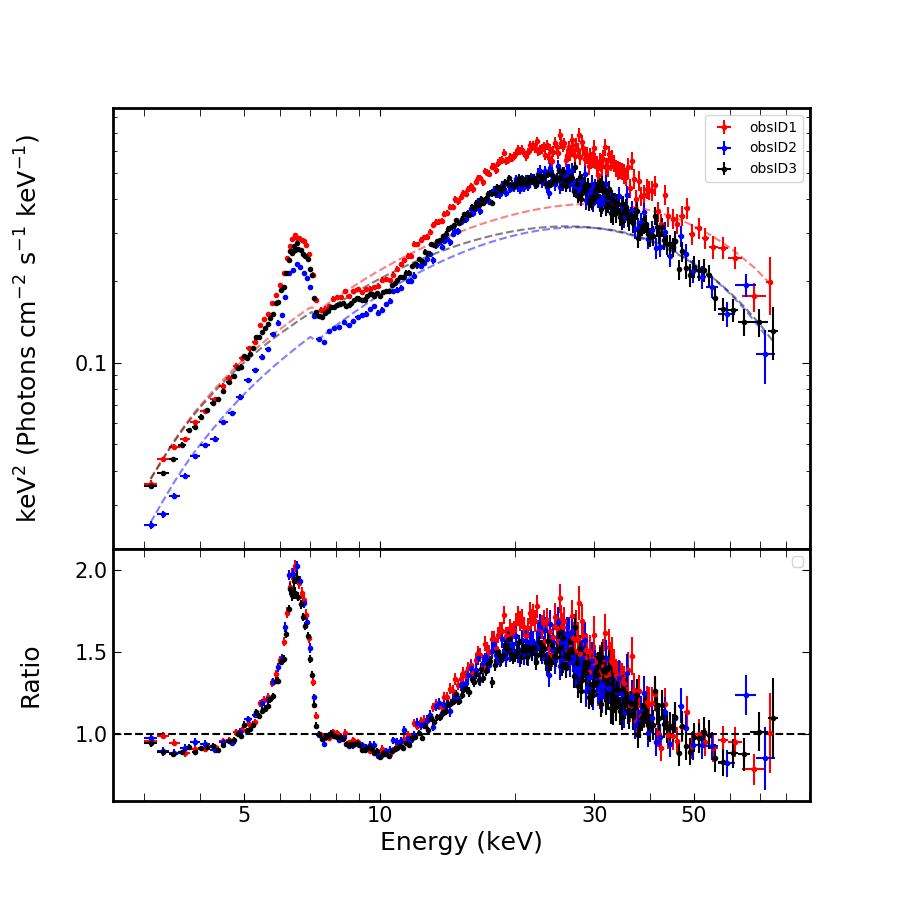}}
\caption{The unfolded \textit{NuSTAR} spectra of obsID1 (red), obsID2 (blue), obsID3 (black) modelled with cutoffpl and galactic absorption for plotting purposes. The fitted models are shown by dotted lines in their respective colours. The plots shows only FPMA data.}
\label{nustar_spec}
\end{figure}

\begin{figure}[ht!]
\centering
\resizebox{\hsize}{!}{\includegraphics[width=1.5\hsize]{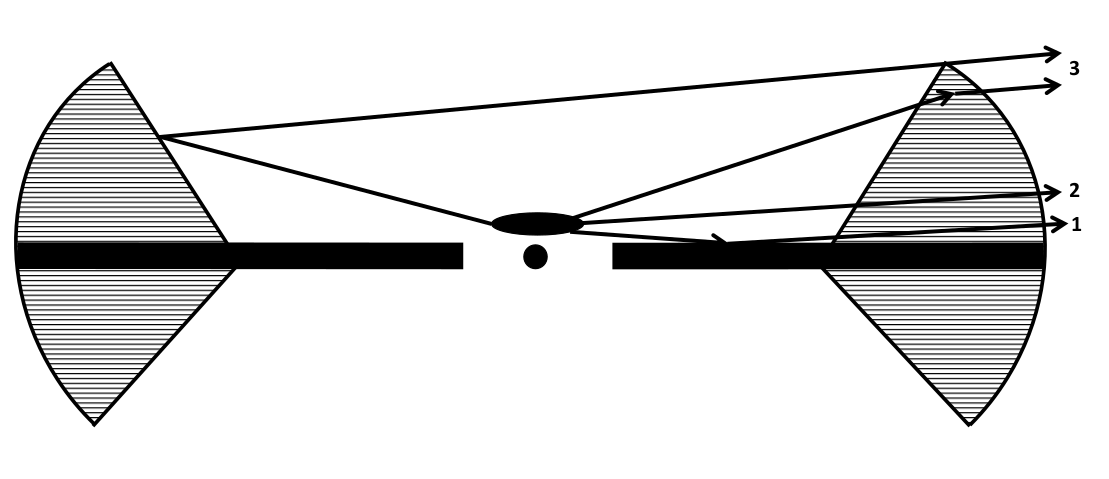}}
\caption{The geometry assumed for the best fit spectral model as outlined in the text. 1, 2, and 3 represents continuum emission, reflection from disc and reflection from wind. The continuum is assumed to originate from the Corona above the black hole. The shaded portion indicate the disc wind, which the black horizontal structure represents the disc.}
\label{spec_model}
\end{figure}

%\begin{table}
%\caption{List of observations}
%\label{table_obs}
%\centering   
%\renewcommand{\arraystretch}{1.5}
%\begin{tabular}{c c c} 
%\hline\hline  
%Observation ID & Date of observation & Exposure \\
%30302020006 & 2018-04-09 & 11.8 ks \\
%30302020008 & 2018-04-19 & 12.9 ks \\
%80401312002 & 2018-06-08 & 26.2 ks \\
%90501321002 & 2019-05-05 & 28.7 ks \\
%30502008002 & 2019-05-19 & 25.7 ks\\
%30502008004 & 2019-07-31 & 23.2 ks\\
%30502008006 & 2019-09-13 & 23.7 ks\\\
%90501346002 & 2019-10-16 & 42.6 ks \\
%\hline
%\end{tabular}
%\end{table}

\section{Spectral Analysis}
The spectral fitting was performed using XSPEC version 12.10.1 \citep{XSPEC1_1996ASPC..101...17A,XSPEC2_2003HEAD....7.2210D}. The spectra from both FPMA and FPMB were grouped using the \textit{grppha} module of the FTOOLS \citep{ftools_2014ascl.soft08004N} in such a way that each bin contains a minimum of 25 counts. As an initial probe, we modelled the spectra in 3-79 keV range in all observations using a cutoff power-law (cutoffpl), along with galactic absorption \citep[tbabs:][]{TBABS_2000ApJ...542..914W} and a constant (const) to account for uncertainties between FPMA and FPMB. Since the goal of the initial test was to model the continuum, we excluded the reflection dominated energy ranges of $5$-$7.5$ keV and $12$-$40$ keV. The neutral hydrogen column density ($N_{\rm H}$) was fixed at $5.0 \times 10^{22} \ \rm cm^{-2}$ as previously seen in this source \citep{Zoghbi_2016ApJ...833..165Z,Koljonen_2020A&A...639A..13K}. We use the $\chi^2$ statistics to determine the goodness of the fit throughout this work. For the initial fit, we got a $\chi^{2}$/dof of 1560/481, 1277/443 and 3070/545 for obsID1, obsID2 and obsID3. We now use the estimated continuum to plot the spectra and residuals in the 3 $-$ 79 keV interval to highlight the Gaussian Fe K$\alpha$ feature at 6.5 keV and a Compton hump at 10 $-$ 50 keV as signatures of reflection often seen in BHBs. The spectrum and the residuals are shown in Fig. \ref{nustar_spec}. The photon index of the fit resulted in unusually low values of 0.81, 0.61 and 0.83 for obsID1, obsID2, and obsID3 suggesting that the spectra are reflection dominated and the continuum is highly absorbed. With different combinations of relativistic reflection models and ionised absorbers, we did not get a reasonable fit. This could be due to the variability in the spectral parameters along with the variability seen as flares in the lightcurves. Hence, we perform a flux resolved spectroscopic analysis as outlined in the next subsections.

\begin{figure}[ht]
\centering
\resizebox{\hsize}{!}{\includegraphics[width=1.5\hsize]{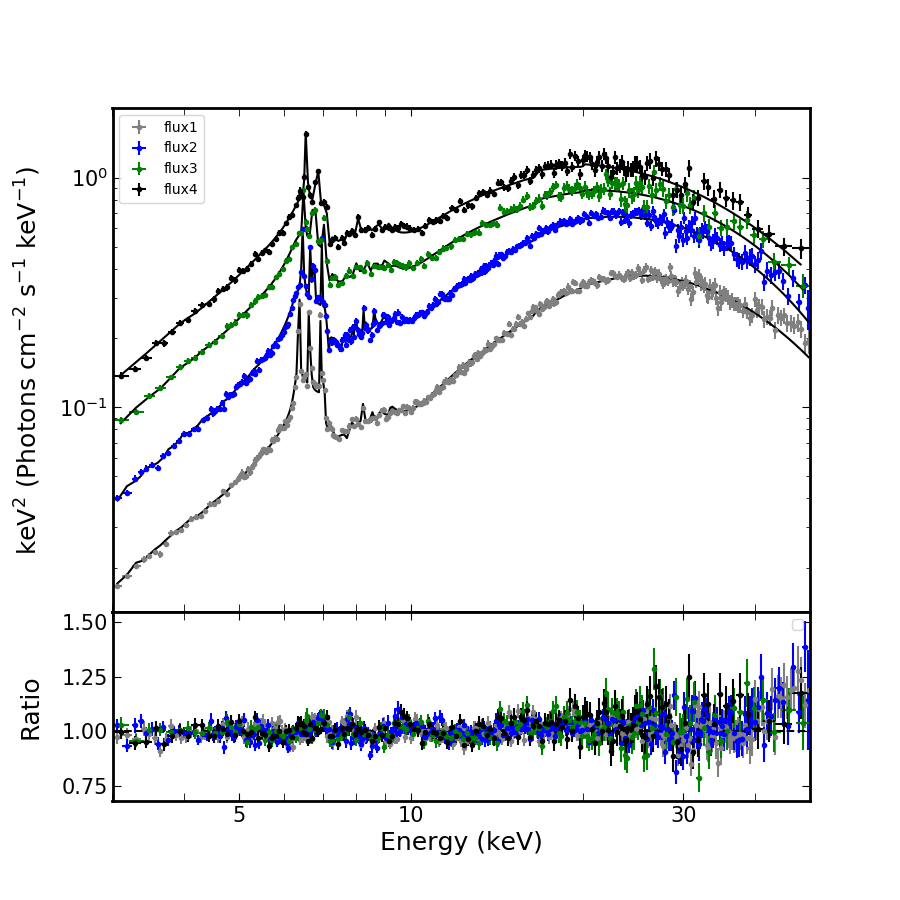}}
\caption{Unfolded \textit{NuSTAR} spectra for different flux levels as outlined in section 3.2. Different colours represent spectra from different flux intervals. The plots shows only FPMA data.}
\label{M2_spec}
\end{figure}

\begin{figure}[h]
\centering
\resizebox{\hsize}{!}{\includegraphics[width=1.5\hsize]{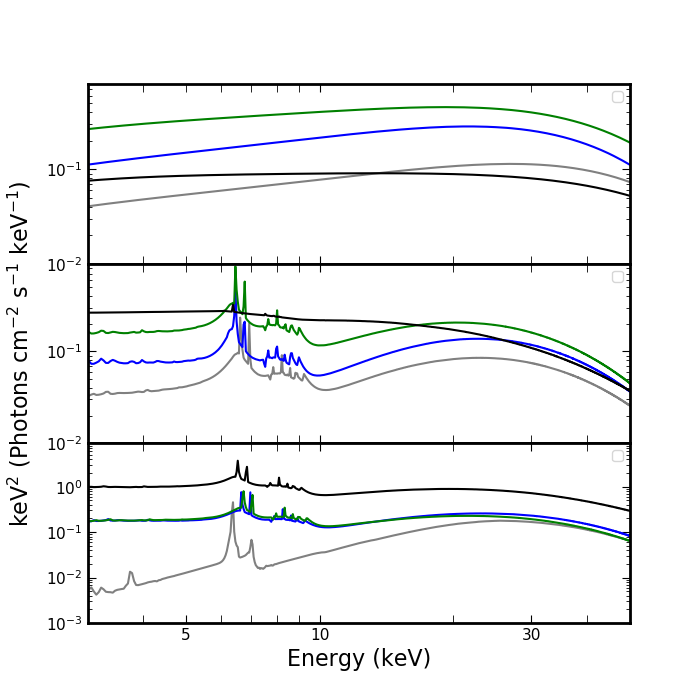}}
\caption{Unabsorbed spectral components at different flux levels. Top: nthcomp, middle: xillverCP2 - wind, and bottom: xillverCP1 - disc. Colours grey, blue, green and black represent flux1, flux2, flux3 and flux4.} 
\label{M2_spec_components}
\end{figure}

%From method 1 we find that the spectra in all low flux and high flux intervals of all three observations have similarities even though certain parameters are not consistent across observations. This inconsistency could be a time interval selection effect as we see in method 2 (see 3.2) that the spectra are variable with respect to the flux. For the continuum we find that the photon index gets softer for the high flux intervals in comparison to the low flux intervals in obsID1 and obsID3 while it remains constant in obsID2. For the ionised absorber we find that the N$_H$ is in the range $30$ to $80$ $\times$ 10$^{22}$ cm$^{-2}$, while the ionisation parameter is in the range of 10 to 100 erg~s$^{-1}$~cm. In all the observations, the wind reflector is highly ionised with ionisation parameter in the range of $10^3$ to $10^5$ erg~s$^{-1}$~cm. However the disk reflector is slightly less ionised in comparison to the wind reflector in most of the cases.

\begin{table*}
\caption{Parameters of the spectral fit from flux resolved spectroscopy.}
\label{table_fluxresolved}
\centering   
\renewcommand{\arraystretch}{2.0}
\begin{tabular}{c c c c c c } 
\hline\hline  
Parameter & Unit & flux1  & flux2  & flux3 & flux4    \\
\hline\hline
flux & ergs cm$^{-2}$ s$^{-1}$ & 8.1 & 15.8 & 22.9 & 30.6 \\
%Parameter & Unit & \multicolumn{2}{c}{obsID1}   & \multicolumn{2}{c}{obsID2}   & \multicolumn{2}{c}{obsID3}   \\
% &  & low flux  & high flux  & low flux  & high flux  & low flux  & high flux\\
\hline
zxipcf  & & & & &  \\ 
\hline 
N$_H$ & $\times10^{22}$~cm$^{-2}$   &  $42^{+1}_{-1}$ & $41^{+1}_{-1}$ & $40^{+2}_{-2}$ & $44^{+1}_{-2}$\\ 
[1ex] 

log($\xi$) & erg~s$^{-1}$~cm & $1.62^{+0.05}_{-0.03}$ & $1.33^{+0.04}_{-0.05}$ & $1.4^{+0.07}_{-0.06}$ & $1.17^{+0.09}_{-0.1}$\\ 
[1ex] 

Redshift &   & $0.005^{+0.001}_{-0.002}$ & $-0.0^{+0.001}_{-0.002}$ & $-0.001^{+0.006}_{-0.002}$ & $-0.003^{+0.003}_{-0.004}$\\ 
[1ex] 

\hline 
nthcomp  & & & & &  \\ 
\hline 
Gamma &   & $1.597^{+0.004}_{-0.011}$ & $1.604^{+0.007}_{-0.006}$ & $1.753^{+0.017}_{-0.02}$ & $1.932^{+0.028}_{-0.031}$\\ 
[1ex] 

kTe & keV & $10.3^{+0.1}_{-0.1}$ & $8.2^{+0.1}_{-0.1}$ & $9.2^{+0.5}_{-0.5}$ & $13.8^{+0.6}_{-0.6}$\\ 
[1ex] 

norm & $\times10^{-3}$ & $15.7^{+0.2}_{-0.3}$ & $43^{+1}_{-3}$ & $111^{+8}_{-12}$ & $35^{+33}_{-26}$\\ 
[1ex] 

\hline 
xillverCP1 - disc  & & & & &  \\ 
\hline 
log ($\xi$) & erg~s$^{-1}$~cm & $2.0^{+0.01}_{-0.04}$ & $3.3^{+0.01}_{-0.02}$ & $3.33^{+0.07}_{-0.04}$ & $3.45^{+0.03}_{-0.03}$\\ 
[1ex] 

z &   & $0.008^{+0.001}_{-0.001}$ & $0.0^{+0.002}_{-0.001}$ & $-0.01^{+0.006}_{-0.006}$ & $0.02^{+0.002}_{-0.002}$\\ 
[1ex] 

norm & $\times10^{-3}$ & $2.4^{+0.04}_{-0.1}$ & $3.9^{+0.2}_{-0.2}$ & $3.9^{+0.4}_{-0.4}$ & $20^{+1}_{-1}$\\ 
[1ex] 

$R_{soft}$ & 2-8 keV & $0.3$ & $1.5$ & $0.7$ & $13.8$\\ 
[1ex] 

$R_{hard}$ & 10-50 keV & $1.2$ & $0.8$ & $0.4$ & $8.7$\\
[1ex] 

\hline 
xillverCP2 - wind  & & & & &  \\ 
\hline 
log ($\xi$) & erg~s$^{-1}$~cm & $3.29^{+0.01}_{-0.02}$ & $3.12^{+0.01}_{-0.02}$ & $3.34^{+0.03}_{-0.03}$ & $>4.33$\\ 
[1ex] 

z &   & $0.005^{+0.001}_{-0.001}$ & $0.032^{+0.001}_{-0.001}$ & $0.03^{+0.002}_{-0.004}$ & $0.089^{+0.029}_{-0.023}$\\ 
[1ex] 

norm & $\times10^{-3}$ & $0.56^{+0.01}_{-0.01}$ & $1.16^{+0.02}_{-0.02}$ & $2.0^{+0.1}_{-0.1}$ & $1.9^{+0.2}_{-0.1}$\\ 
[1ex] 

$R_{soft}$ & 2-8 keV & $0.9$ & $0.7$ & $0.6$ & $3.3$\\
[1ex] 

$R_{hard}$ & 10-50 keV & $0.6$ & $0.4$ & $0.4$ & $1.8$\\ 
[1ex]

\hline
$\chi^2$/d.o.f & & $2237/1905$ & $2069/1797$ & $1529/1392$ & $1428/1357$ \\

\hline
\hline
\end{tabular}
\end{table*}

\begin{figure*}[ht!]
\centering
\begin{subfigure}[b]{0.4\textwidth}
\centering
\resizebox{\hsize}{!}{\includegraphics[width=\textwidth]{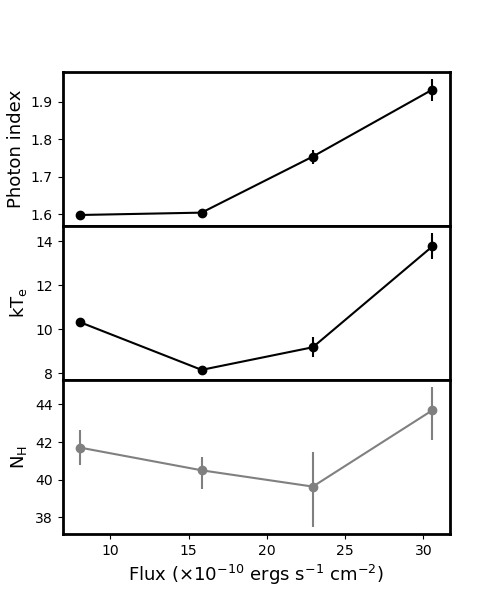}}
%\caption{}
\label{M2_spec_par_1}
\end{subfigure}
\begin{subfigure}[b]{0.4\textwidth}
\centering
\resizebox{\hsize}{!}{\includegraphics[width=\textwidth]{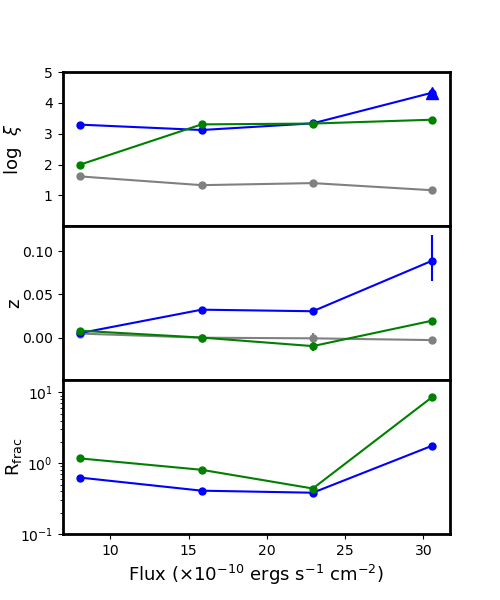}}
%\caption{}
\label{M2_spec_par_2}
\end{subfigure}

\caption{The spectral parameters from the flux resolved spectroscopy with respect to the flux. Colours black, grey, blue and green represent the spectral parameters of continuum (nthcomp), ionised absorber (zxipcf), wind reflector (xillverCP1) and disk reflector (xillverCP2). Left: Top, middle and bottom shows photon index, electron temperature and N$_H$ of the ionised absorber. Right: Top, middle and bottom shows ionisation parameter, red-shift and $R_{hard}$}
\label{M2_spec_par}
\end{figure*}

\subsection*{Flux resolved spectroscopy}
We divide the three observations into four different flux levels, and combine similar flux levels from different observations. The main aim of this method is to explore if there is a change in any of the spectral parameters with respect to the flux. If there is a change either the spectral parameters or the geometry of absorber and reflector, then we can expect a respective change in polarisation degree and angle. The time intervals for flux resolved spectra were selected from a 10 seconds binned lightcurve of FPMA. The flux levels correspond to a count-rate intervals of  $<$15 c s$^{-1}$, 15-30 c s$^{-1}$, 30-40 c s$^{-1}$ , and $>$40 c s$^{-1}$. This corresponds to an average flux of 8.07, 15.83, 22.93, and 30.56 $\times 10^{-10}$ ergs cm$^{-2}$ s$^{-1}$ in 3 $-$ 50 keV range. We generate GTIs for different flux levels and re-generate the spectra in those time intervals using \textit{nuproducts}. We combine the respective spectra using the addspec tool (version 1.3.0) of FTOOLS. Before combining the fit, we also verified that the spectra from same flux levels from different observations are consistent with each other.

We modelled all the spectra using a Comptonisation model \citep[nthcomp:][]{nthcomp1_1996MNRAS.283..193Z,nthcomp2_1999MNRAS.309..561Z}, two reflection models for the reflection from the accretion disc and the obscurer \citep[xillverCP][]{xillver1_2010ApJ...718..695G,xillver2_2011ApJ...731..131G,xillver3_2013ApJ...768..146G}, galactic absorption \citep[tbabs:][]{TBABS_2000ApJ...542..914W}, a multiplicative ionised absorber \citep[zxipcf:][]{zxipcf1_2008MNRAS.385L.108R,zxipcf2_2006A&A...453L..13M} on the direct continuum and on the disc reflector. The neutral absorption $N_{\rm H}$ was frozen at 5.0 $\times$ 10$^{22}$ cm$^{-2}$ for all the observations and a constant was used to account for the uncertainties in the calibration between FPMA and FPMB as mentioned in the initial fit. Hence the final model in XSPEC notation is const $\times$ tbabs $\times$ (zxipcf $\times$ (nthcomp+xillverCP1) + xillverCP2). Here we assume the xillverCP1 to be the disc component and xillverCP2 to be the wind. The assumed geometry of the spectral model is shown in Fig. \ref{spec_model}. Here the reflected component from the wind could either be from the side towards line of sight or on the other side. We find an excess above 50 keV, may be connected with a putative jet emission \citep{Kong_2021ApJ...906L...2K}. We do not focus on it's origin here and it does not affect our spectral modelling, which is dominated by the spectrum below 20 keV. The covering fraction in zxipcf was frozen to 1.0 as it was found to be always consistent with unity. We used a disc black body as a seed in the nthcomp Comptonisation model. Since the spectra in the \textit{NuSTAR} energy range are not very sensitive to the seed black body temperature, we fix it at 0.8 keV which is a reasonable value for the hard state of GRS 1915+105. xillverCP was used to model the reflection component only, and hence we tied the electron temperature and photon index within the xillverCP model to the same parameters in nthcomp. However, to compute the reflected spectrum, xillverCp uses a nthcomp model with seed black body temperature of 0.05 keV, which might cause some uncertainties in model especially in soft X-ray below 5 keV. When left free the iron abundance (A$_{Fe}$) was constrained between 2.0 and 3.0 and we fixed A$_{Fe}$ to 2.5. Super solar iron abundances were seen previously in this source \citep{Lee_2002ApJ...567.1102L, Ueda2009_2009ApJ...695..888U, Kong_2021ApJ...906L...2K}. The inclination angle of the system has been estimated to be 66$^{\circ}$ $\pm$ 2$^{\circ}$ \citep{Fender_1999MNRAS.304..865F}, so we fixed the disc inclination to this value. We also fixed the wind inclination angle to 30$^{\circ}$, since this parameter was insensitive to the fit. The 'refl\_frac' in xillver defined as the ratio of flux incident on the disk to the flux escaping into the line of sight \citep{Dauser_2016A&A...590A..76D}, was also frozen to 1 as this parameter was poorly constrained in the fit. We also calculated the ratio of the flux of reflected components to that of the continuum
in 2-8 keV ($R_{soft}$) and 10-50 keV ($R_{hard}$) energy intervals, to understand if the intrinsic flux is dominated by the reflected component in comparison to the direct continuum. The parameters of the fit are outlined in Table. \ref{table_fluxresolved} and Fig. \ref{M2_spec} shows the modelled spectra of all observations. Fig. \ref{M2_spec_components} shows the unabsorbed continuum and both the reflectors at different flux levels. \\

%We used the same model and the same frozen parameters used in method 1 for the spectral fit. As done for method 1, we again calculate the unabsorbed reflected fraction for both the wind and the disk components in comparison to the continuum in 2-8 keV and 10-50 keV. The parameters of the fit are outlined in Table. \ref{table_fluxresolved} and Fig. \ref{M2_spec} modelled spectra from method 2. Fig. \ref{M2_spec_components} shows the unabsorbed continuum and both the reflectors at different flux levels. \\

The parameters from the spectral fit are plotted with respect to the flux in Fig. \ref{M2_spec_par}. The main finding is that the photon index is correlated to the flux. The electron temperature is also seen to increase after a drop from the first to the second flux level. This indicates that the flux change is correlated to changes in the central engine. We find that the properties of the ionised absorber do not change with respect to the flux (see Fig. \ref{M2_spec_par}). The ionisation parameters of the disk and wind reflectors are in the range of 10$^2$ to 10$^4$ erg~s$^{-1}$~cm and 10$^3$ to 10$^5$ erg~s$^{-1}$~cm, respectively. We also see an increase in the red-shift with respect to the flux for the wind reflector, while no clear trend is seen for the red-shift of the disc component. Other than the highest flux level the $R_{hard}$ is constant, while in the highest flux level, $R_{hard}$ increased by an order of magnitude for the disk and by a factor of 5 for the wind. \\

These spectral models and results from our analysis are similar to the results from \cite{Koljonen_2020A&A...639A..13K} in their epoch 3.
%However since the continuum spectra is changing with respect to the flares, we got a better fit \textbf{in comparison to their best fit $\chi^2$/dof values of 666/441 and 716/439 for their model A and model B.} 
However, from our analysis it is evident that the continuum as well as the reflectors are changing with respect to the flux. Since two distant reflection components (xillverCP) could model the observed spectra, it is evident that the reflection is not from the inner disk. However one of the reflection components could be from a relatively distant disk component. At least in some flux levels, the spectrum required a positive red-shift for the assumed disc component. This could be because reflection from the receding side of the disk is more prominent because the one from the approaching side is more absorbed. The wind reflector is always red-shifted suggesting reflection from in-falling matter which is further consistent with the claim of reflection from a failed disk wind by \cite{Miller_2020arXiv200707005M}. It could also be possible that this reflected component is from the wind outflow observed from the other side of the source. There is also a gradual increase seen in the red-shift of the wind with respect to the flux suggesting that the outflow is faster with respect to the flux. In the framework of MHD disk winds, an increase in speed is expected following an increase in flux for sources with relatively high Eddington ratios \citep{Fukumura_2018}. This is due to the fact that the increase in mass accretion rate, which drives an increase in flux, at the same time provides a significant increase in the density at the base of the disk wind. Therefore, the ionisation front of the wind can move inward and we would observe the same ionic species being ejected from smaller radii, thus with larger speeds. The variability time scales of few tens of seconds of the flares are comparable to the viscous time scales at a radius of 100 $GM/c^2$ \citep{Balakrishnan_2020arXiv201215033B}. We see that the $\xi$, red-shift, and N$_H$ of the absorber are rather constant again indicating that the flares in the lightcurve are not due to the change in the column density of the absorber. There is no clear trend with respect to the flux for both reflectors and we find that the $\xi$ of the reflectors ranges from $10^{2}$ to $10^{5}$ erg~s$^{-1}$~cm. The $\xi$ and $N_H$ inferred from our analysis are also similar to the ones found by \cite{Miller_2020arXiv200707005M,Balakrishnan_2020arXiv201215033B}, in the Compton thick state. Even though \cite{Koljonen_2020A&A...639A..13K,Miller_2020arXiv200707005M, Balakrishnan_2020arXiv201215033B} use a neutral intrinsic absorption, we get a better fit using a ionised absorber model (zxipcf) with moderately ionised medium. \\

The Eddington ratio calculated in the 0.1 to 500 keV band considering only the continuum emission for the lowest flux range is approximately 0.03, similar to what is seen in \cite{Balakrishnan_2020arXiv201215033B} for some observations. This indicates that the source is not going into quiescence state but could be an intrinsic new state. Our results also indicate that the recent drop in flux is partially due to absorption as well as a diminished central engine, as also claimed by \cite{Koljonen_2020A&A...639A..13K,Miller_2020arXiv200707005M}. The photon index of the source is in the range 1.5 to 2.0 which reveals that the source is harder than before. As MHD winds can be possible in this system very close to the central engine \citep{Ratheesh_2021}, a magnetically driven accretion disk wind becomes a strong candidate for the reflection and absorption. For low values of accretion rate, the magnetic field may fail in accelerating the matter and hence form a failed disk wind \citep{Miller_2020arXiv200707005M}, which is consistent with the red-shift seen in the wind reflection from our analysis. From a similar flux resolved spectroscopy of a flare in \textit{NICER} data, \citep{Neilsen_2020ApJ...902..152N} finds two component intrinsic absorption from the outer disk. However an increase in the red-shift of the reflector along with a change in the Coronal properties might indicate that the reflector is closer to the central engine at-least in our spectral models. However, if the reflection is also from this outer disk, difference in polarimetric signatures would constrain the obscurer geometry. The limitation of the spectral analysis is that the reflector and absorber geometry is not constrained. Also it is not clear if there is a change in the geometry of the reflector with respect to the continuum.\\

\section{Polarisation properties of the reflecting matter}
Even though the spectral results showed the presence of a reflector and absorber in this particular state of GRS 1915+105, the geometry of the reflector remains unconstrained by spectroscopy alone. In this section we calculate the polarisation properties of the reflecting matter, using the radiative transfer Monte Carlo code described in \cite{Ghisellini_1994MNRAS.267..743G} (hereinafter GHM94).

\begin{figure}[hbt!]
\centering
\resizebox{\hsize}{!}{\includegraphics[width=1.5\hsize]{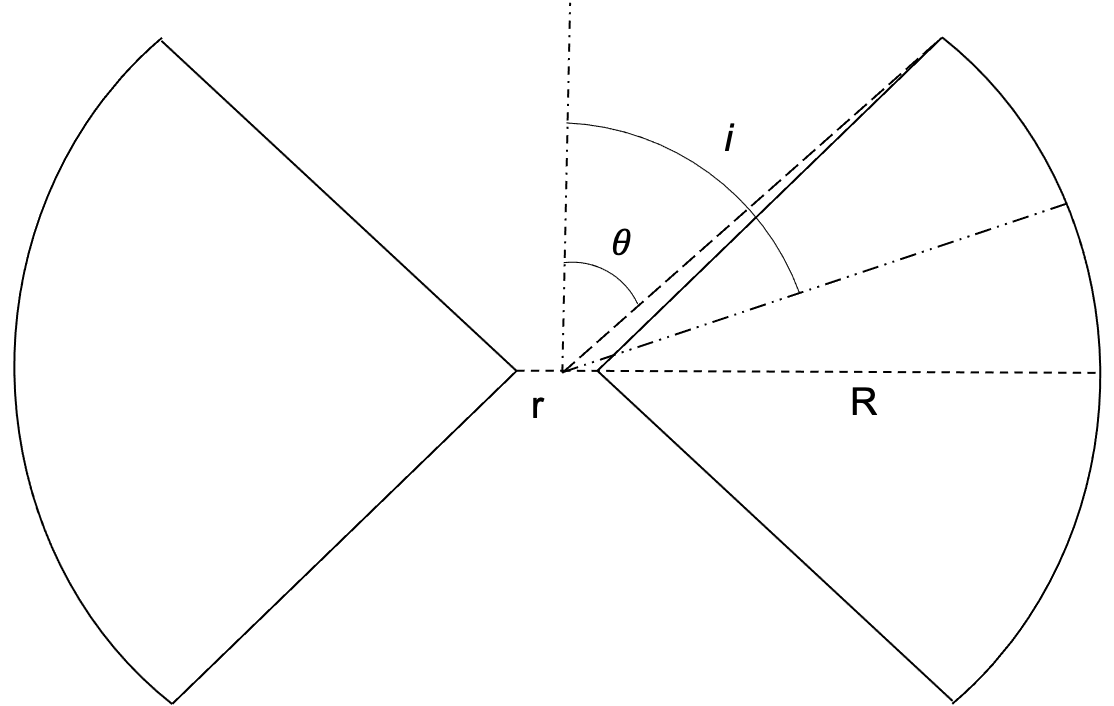}}
\caption{The geometry used for the polarimetric simulation as outlined in the text. $i$ is the inclination angle, $\theta$ is the half opening angle, $r$ and $R$ are the inner and outer radius of the outflow.  }
\label{sim_raydiagram}
\end{figure}

While the details of the code, and in particular the atomic data used, can be found in GHM94 and expecially in \cite{Matt91_1991A&A...247...25M}, here we recall its main characteristics. Neutral, cold matter is assumed, with solar chemical abundances. Interactions between photons and matter which are considered in the code are Compton scattering and photoelectric absorption. If absorption occurs in the K-shell of an iron atom, fluorescent emission (both K$\alpha$ and K$\beta$) is also included, with a fluorescent yield of 0.34 (fluorescent photons are assumed to be unpolarised).  The geometry of the matter is illustrated in Fig.\ref{sim_raydiagram} (see also Fig.1 of GHM94), with a ratio between inner and outer radii of the matter (hereinafter R) fixed to 0.1 or 0.5. We explored different values of $\theta$, the half-opening angle of the matter configuration, of the equatorial hydrogen column density N$_{H,refl}$, and two opposite scenarios for the matter ionisation: neutral matter and fully ionised matter. In the latter case, the only interaction between photons and matter is the Compton scattering  (in practice, the photoabsorption cross section is set to zero). With the different scenarios mentioned above, we calculate the polarisation degree in the 2-10 keV energy band overlapping the 2-8 keV energy band of IXPE.  The input spectrum is a power law in the 2-30 keV energy range, with a photon index of 2. While the photon index of the source is found, in some state, to be flatter than the value adopted for the simulations, we verified that the emerging polarisation is largely insensitive to this parameter. We note that the adoption of either a neutral of a fully ionised matter is an oversimplification of the real situation, in which the matter is often found to be ionised but not fully so. The present version of the code does not allow us to deal with partial ionisation, but we believe that our simplified assumptions are good enough to highlight the importance of polarisation in constraining the geometry of the reflectors. More detailed calculations, allowing also for partial ionisation of the matter, are then deferred to a future paper.\\

Recently, many Monte Carlo codes have been developed which treat polarisation properties of Comptonisation \citep[e.g. ][]{Schnittman_Krolik_2010ApJ...712..908S,Tamborra_2018A&A...619A.105T, Zhang_2019ApJ...875..148Z}. These codes assume for the Comptonising medium a pure electron cloud, and therefore are not adequate for the astrophysical situation we are studying. To our knowledge, the only other code which can in principle deal with the same problem is STOKES \citep{Marin_2018MNRAS.478..950M,Marin_2018MNRAS.473.1286M,Goosmann_Gaskell_2007A&A...465..129G}. However, the publicly available version of the code is tailored for IR-Optical-UV radiation, rather than X-rays, so we decided at this stage to use the GHM94 code. \\

In the next two paragraphs, we will present the cases of neutral and fully ionized matter assuming unpolarised
illuminating radiation. Then, we will discuss the case of polarised primary radiation.\\

\subsection{Neutral matter}
We first discuss the results for neutral matter case.
In Fig.\ref{poldegCosi_n} the 2-10 keV polarisation degree as a function of the cosine of the inclination angle $i$ of the system is shown for different values of N$_{H,refl}$ and $\theta$. The primary spectrum, assumed unpolarised, is a power law with $\Gamma$=2 in the energy range 2-30 keV. The emerging photons are then grouped in 20 angular bins, equally spaced in cos $i$. The polarisation is found to be always perpendicular to the symmetry axis of the system. For $i$ lower than $\theta$ the polarisation degree is low because of the dilution of the primary emission, which is unobstructed for those values. The polarisation degree is also low for small values of the equatorial column density, because in those cases the matter is partially transparent, with consequent dilution by the primary radiation which pierce through. This effect is also clear in Fig.\ref{poldegEne_n}, where the dependence of the polarisation degree on the energy is shown, as an illustration, for an inclination angle of 65$^{\circ}$ (consistent with the one of GRS 1915+105), for four different $\theta$ values (30$^{\circ}$, 45$^{\circ}$, 60$^{\circ}$ and 75$^{\circ}$), three different N$_{H,refl}$ values (1, 3, and 5 $\times$ $10^{24}$ cm$^{-2}$) and two different values of R (0.1 and 0.5).\\

A strong decrease of the polarisation degree with energy is observed for low values of N$_{H,refl}$, when the matter becomes transparent above a few keV. A much more constant polarisation degree is instead observed for high values of N$_{H,refl}$, when the matter is always optically thick. The two drops in the polarisation degree are due to the iron K$\alpha$ and K$\beta$ emission lines because fluorescent line emission is unpolarised.
\subsection{Fully ionised matter}
In the case of neutral matter, most of the reflected radiation is scattered once or a few times at most. When the matter is fully ionised, however, and if it is optically thick, photons can suffer many scatterings before escaping. As a result, the photon distribution is isotropic and the polarisation degree is much lower. This can be seen in Figs.\ref{poldegCosi_i} and \ref{poldegEne_i}, which are the same as Figs.\ref{poldegCosi_n} and  \ref{poldegEne_n} but in the fully ionised scenario. In this case, the polarisation degree is well below 10\% except the case for a $\theta$ of $75^{\circ}$, and almost energy-independent because at these energies the scattering cross-section is still almost constant at the Thomson value. The modest increase in polarisation degree for large optical depths is due to the fact that, because of Compton down-scattering, low energy photons emerging from the matter are on average scattered more times than high energy photons. The polarisation angle is again perpendicular to the symmetry axis of the system.

\begin{figure}[hbt!]
\centering
\resizebox{\hsize}{!}{\includegraphics[width=1.5\hsize]{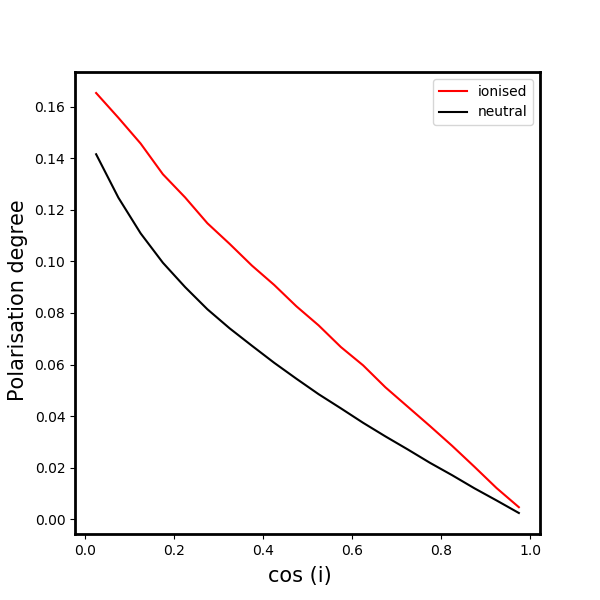}}
\caption{Polarisation degree with respect to different inclination angles for fully ionised and neutral scenario of a  simple disk geometry.}
\label{poldeg_incl_disk}
\end{figure}
\begin{figure*}[hbt!]
\centering
\begin{subfigure}[b]{0.45\textwidth}
\centering
\resizebox{\hsize}{!}{\includegraphics[width=\textwidth,height=0.4\textheight]{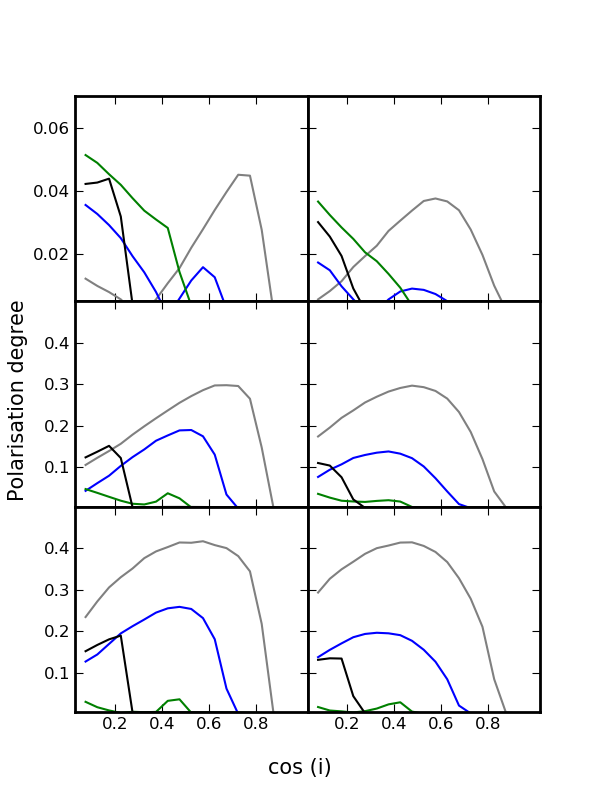}}
\caption{Neutral matter}
\label{poldegCosi_n}
\end{subfigure}
\begin{subfigure}[b]{0.45\textwidth}
\centering
\resizebox{\hsize}{!}{\includegraphics[width=\textwidth,height=0.4\textheight]{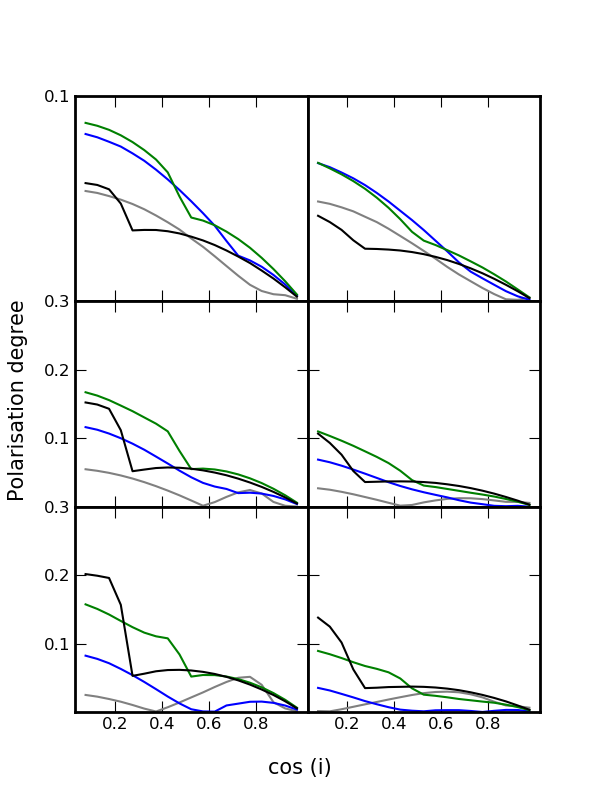}}
\caption{Ionised matter}
\label{poldegCosi_i}
\end{subfigure}
\begin{subfigure}[b]{0.45\textwidth}
\centering
\resizebox{\hsize}{!}{\includegraphics[width=\textwidth,height=0.4\textheight]{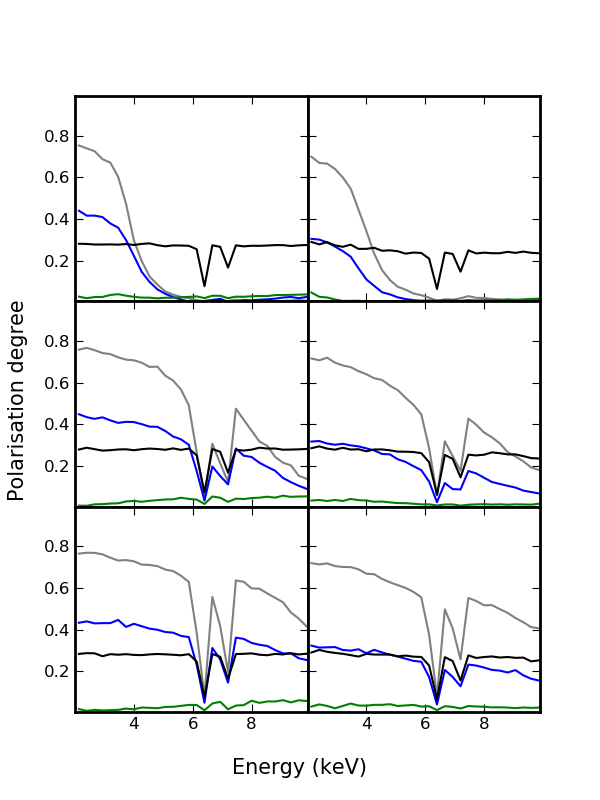}}
\caption{Neutral matter}
\label{poldegEne_n}
\end{subfigure}
\begin{subfigure}[b]{0.45\textwidth}
\centering
\resizebox{\hsize}{!}{\includegraphics[width=\textwidth,height=0.4\textheight]{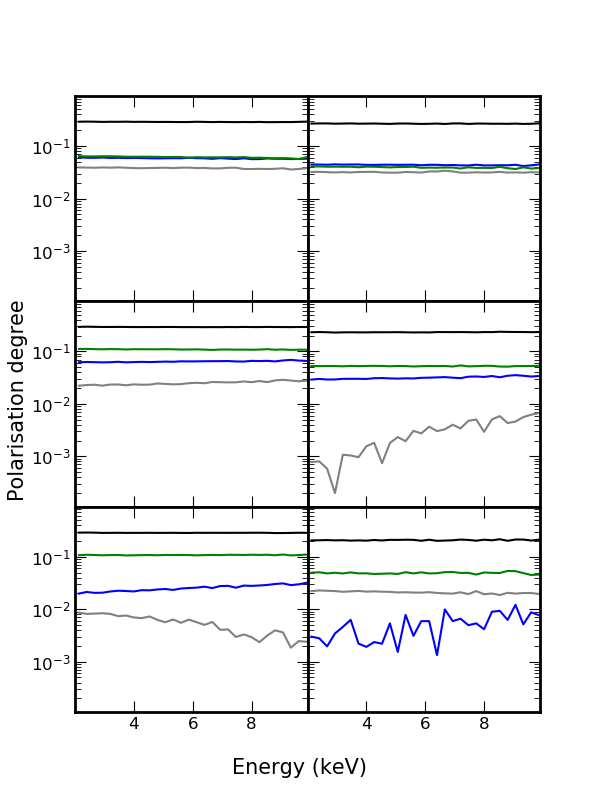}}
\caption{Ionised matter}
\label{poldegEne_i}
\end{subfigure}
\caption{The polarisation degree as a function of cosine of the inclination angle and energy. Panels (a) and (c) refer to neutral matter, instead panels (b) and (d) refer to ionised matter. In each of the plots, the left and the right represents R = 0.1 and 0.5. Top, middle and bottom represents N$_{H,refl}$ values of 1,3,5 $\times$ $10^{24}$ cm$^{-2}$. Colours of grey, blue, green and black represents opening angles of 30$^{\circ}$, 45$^{\circ}$, 60$^{\circ}$ and 75$^{\circ}$.}
\label{poldeg_incl_Ene_n_i}
\end{figure*}
\begin{figure*}[hbt!]
\centering
\begin{subfigure}[b]{0.45\textwidth}
\centering
\resizebox{\hsize}{!}{\includegraphics[width=\textwidth, height=0.4\textheight]{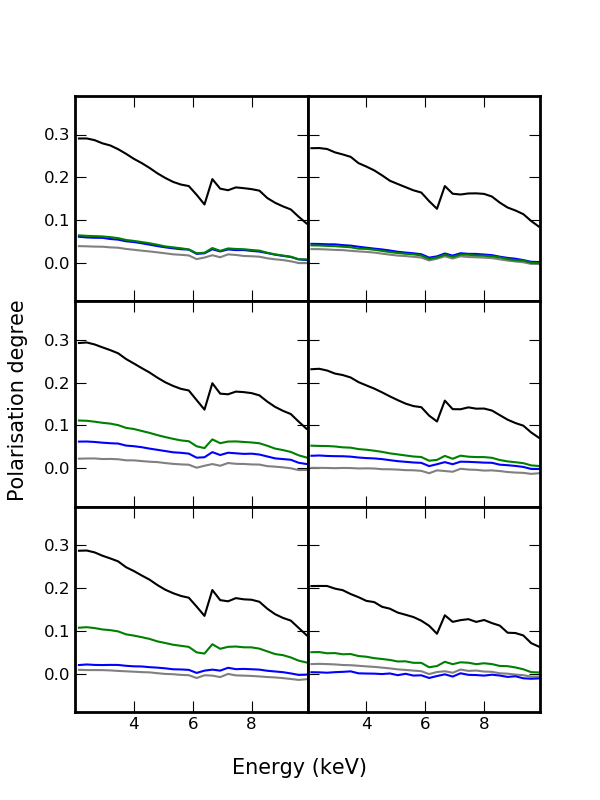}}
\caption{flux1}
\label{poldegEne_flux1}
\end{subfigure}
\begin{subfigure}[b]{0.45\textwidth}
\centering
\resizebox{\hsize}{!}{\includegraphics[width=\textwidth, height=0.4\textheight]{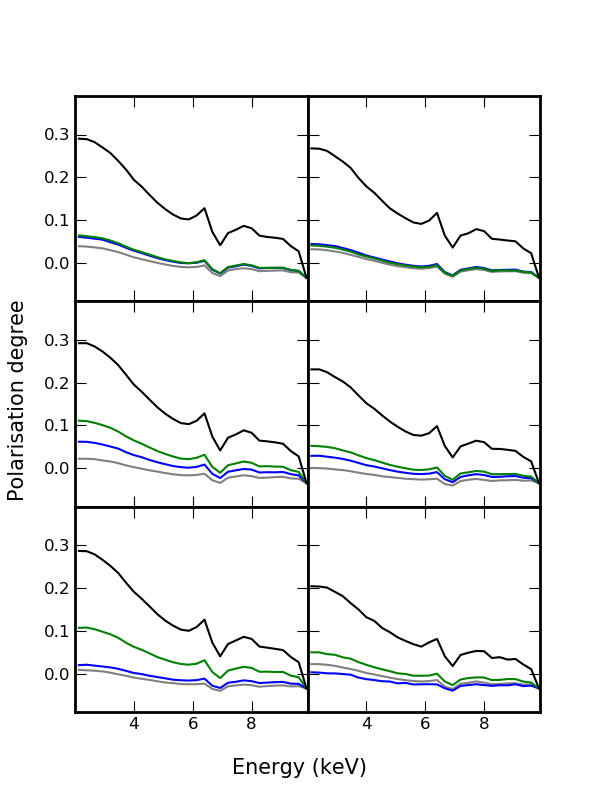}}
\caption{flux2}
\label{poldegEne_flux2}
\end{subfigure}
\begin{subfigure}[b]{0.45\textwidth}
\centering
\resizebox{\hsize}{!}{\includegraphics[width=\textwidth, height=0.4\textheight]{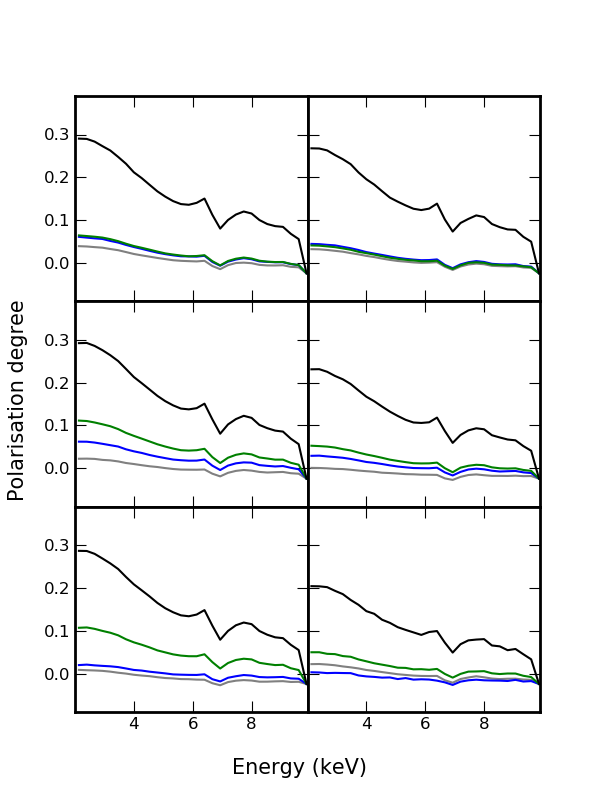}}
\caption{flux3}
\label{poldegEne_flux3}
\end{subfigure}
\begin{subfigure}[b]{0.45\textwidth}
\centering
\resizebox{\hsize}{!}{\includegraphics[width=\textwidth, height=0.4\textheight]{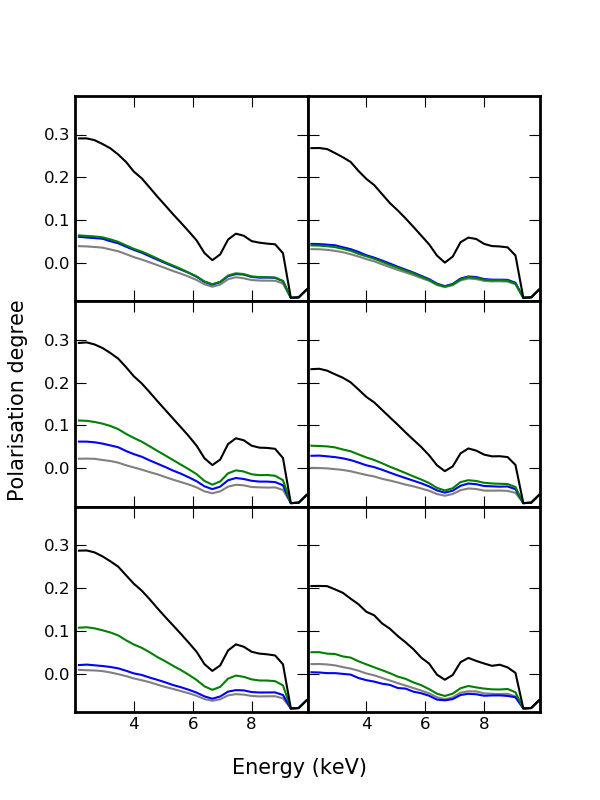}}
\caption{flux4}
\label{poldegEne_flux4}
\end{subfigure}
\caption{The polarisation degree as a function of energy for GRS 1915+105 for different flux levels. Panels (a), (b), (c) and (d) represents flux1, flux2, flux3 and flux4 as outlined in section 3. In each of the plots, the left and the right represents R = 0.1 and 0.5. Top, middle and bottom represents N$_{H,refl}$ values of 1,3,5 $\times$ $10^{24}$ cm$^{-2}$. Colours of grey, blue, green and black represents opening angles of 30$^{\circ}$, 45$^{\circ}$, 60$^{\circ}$ and 75$^{\circ}$.}
\label{grs1915_pol_cal}
\end{figure*}
\subsection{Polarised primary radiation}
 In the previous paragraphs we calculated the polarisation of the radiation reflected by the outflow in the assumption of unpolarised primary radiation. However, if, as widely believed, the primary radiation originates in a hot corona which Comptonises thermal radiation from the disk, a certain level of polarisation is expected, which depends on the geometry of the corona. For instance, \cite{Tamborra_2018A&A...619A.105T} explored two possible geometries for the corona, a slab covering the disk and an hemisphere centred on the black hole. The results depend on the coronal shape, the optical depth of the corona and also the polarisation of the thermal disk emission (the latter was assumed to be either unpolarised or polarised following the Chandrasekhar formula, i.e. with a polarisation, parallel to the disk, ranging from zero for a disk viewed face-on to almost 12\% for an edge-on view). Typically, the coronal polarisation
ranges from zero (for emission normal to the disk) to a maximum of 4\% or so, with a polarisation degree either parallel or perpendicular to the disk axis \citep[see e.g. Fig.16 of ][]{Tamborra_2018A&A...619A.105T}. %https://www.overleaf.com/project/5f0dc8e38c274400019a5cbd

To assess the effect of the polarisation of the primary emission, and remembering that the result for an arbitrary
polarisation degree of the incident radiation can be obtained by properly combining the Stokes parameters of the reflected components arising from unpolarised and 100\% polarised primary radiation, we simulated the response of the wind for a fully polarised primary radiation (independently, for simplicity, of the direction of emission), assuming a polarisation angle either perpendicular or parallel to the disk. \\

The polarisation of the reflected radiation results to be still perpendicular to the disk axis, independently of the polarisation angle of the primary radiation. It may be very large, reaching almost 80-90\% in case of perpendicular
polarisation and neutral matter and/or small optical depths, when the reflection is dominated by photons scattered once, at least in the classical X-ray band. The degree of polarisation diminishes if the matter is ionised and the optical depth is large, because when photons undergo a large number of scatterings before escaping, the initial polarisation is largely forgotten. Indeed, for ionised matter and N$_{H,refl}$=10$^{25}$ cm$^{-2}$, the polarisation of the reflected radiation is only a few percent.

Because the polarisation angle of the reflected radiation is always the same, the total polarisation degree can be estimated, for each energy and inclination angle, by this simple formula: $P \sim P_U*(1-P_{in})+P_P*P_{in}$, where $P_U$ and $P_P$ 
are the polarisation degrees of the reflected radiation illuminated by unpolarised and fully polarised primary emission, respectively, while $P_{in}$ is the actual polarisation degree of the primary radiation (averaged over the emitting angle). Given that the last value is expected to be about 2-3\% \citep{Tamborra_2018A&A...619A.105T}, even in the case of neutral matter and small optical depths (when $P_P$ is the highest) the polarisation degree of the reflecting matter may be only slightly larger than what have been calculated in the previous paragraphs. Given the uncertainties and limitation of our results discussed above, we decided to ignore in the following the small effects of the polarisation of the primary radiation.

\subsection{The case of GRS 1915+105}
We calculate the expected polarisation from the disk and the wind in the different flux levels estimated from the section 3. For the wind, we use the above mentioned scenarios (only from the side towards the line of sight), while for the case of disc we did a similar Monte-Carlo simulation for both neutral and fully ionised slab isotropically illuminated by a point source from above \citep[see ][]{Matt_1989ESASP.296..991M,Matt_1991A&A...247...25M}. The continuum emission coming from the corona is assumed to be unpolarised. In case of the disk, the polarisation angle is parallel to the symmetry axis of the system and perpendicular to the polarisation angle from the wind reflector. Fig. \ref{poldeg_incl_disk} shows the inclination angle dependence of the polarisation degree in case of the disk reflector. We also assume fully ionised state for log $\xi$ greater than 2.5 and neutral for log $\xi$ lesser than 2.5. Hence, in the case of the first flux level, we assume the disk to be neutral while the wind is fully ionised. In the other flux levels, both the disk and wind are fully ionised in these calculations. We used the absorbed flux in each energy bin, to calculate the resulting polarisation degree. We did this calculation for different cases of opening angle, N$_{H,refl}$ and R for the wind as outlined in the above subsections. The results of these calculations are given in Fig. \ref{grs1915_pol_cal}. Since the polarisation from the disk and wind are perpendicular to each other, we assume the wind polarisation to be positive and the disk to be negative. We find that the highest polarisation degree (0.1 to 0.3) is when the opening angle of the wind is 75$^{\circ}$. In the case of the first flux level the wind dominates most of the spectra, while in the other flux levels the disk dominates beyond approximately 6 keV at least for low values of the opening angle.
\section{Discussion and conclusions}
In this work,  we explored how X-ray spectro-polarimetry can be used to constrain the geometrical parameters of the matter in the environment of the black hole in GRS 1915+105 in its present, obscured state. To do that, we first
analysed three \textit{NuSTAR} observations of GRS 1915+105 in the recent low flux state. 
%We found flares in the lightcurves for all the three observations as also reported by other authors %\citep{TOO_Iwakiri_2019ATel12787....1I,TOO_Neilsen_2019ATel12793....1N,TOO_Jithesh_2019ATel12805....1J}. 
Thanks to the wide hard X-ray energy band coverage of \textit{NuSTAR} we were able to constrain the continuum, which is heavily absorbed in the soft X-rays. \textit{NuSTAR} observations indicate reflection dominated spectra (see Sec.\ref{nustar_spec}), suggesting that the central engine is still active and intrinsic absorption might be the reason for the observed low flux. 
%For the spectral analysis, we followed two methods. In the first method we divided all the three observations into two flux levels, based on the time intervals at which the countrate was lesser and greater than the mean from a 10 seconds binned lightcurve. In the second method, we selected four different countrate levels from a 10 seconds binned lightcurve and co-added the spectra from corresponding flux levels from different observations. For the spectra from both methods we obtained the best fit for a model comprising of a continuum (nthcomp), 2 reflectors (xillverCP), and an ionised absorber (zxipcf) on the continuum and one of the reflectors. As the inclination of both reflectors were not constrained, we fix it to 66$\degree$ (assumed to be the disc) and 30$\degree$ (assumed to be the wind) in case of the absorbed and unabsorbed reflector. \\
\begin{figure*}[htb!]
\centering
\begin{subfigure}[b]{0.4\textwidth}
\centering
\resizebox{\hsize}{!}{\includegraphics[width=\textwidth]{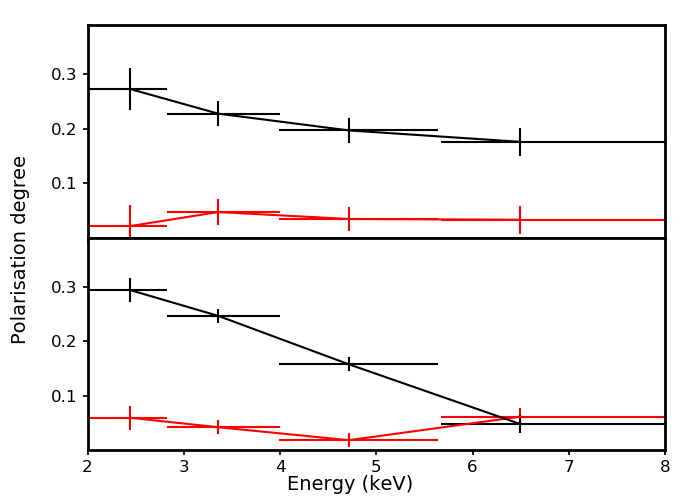}}
%\caption{Polarisation degree}
%\label{ixpeobssim_poldeg}
\end{subfigure}
\begin{subfigure}[b]{0.4\textwidth}
\centering
\resizebox{\hsize}{!}{\includegraphics[width=\textwidth]{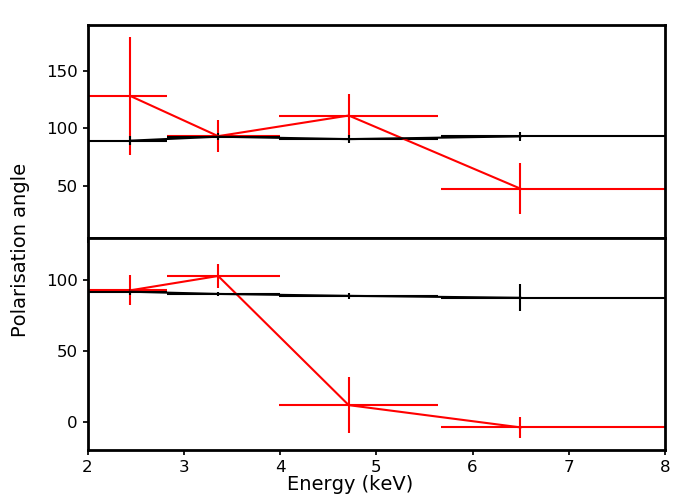}}
%\caption{Polarisation angle}
%\label{ixpeobssim_polang}
\end{subfigure}
\caption{Results of the the IXPE simulations using \textit{ixpeobssim}. The panel on the left refers to the polarisation degree, instead the panel on the right refers to the polarisation angle. The top and bottom are for the lowest flux level and the highest flux levels. The red and black colour indicates a $\theta$ of 30$^{\circ}$ and 75$^{\circ}$.} 
\label{ixpeobssim_fig}
\end{figure*}

The main result of the spectral analysis is that the photon index is changing with respect to the change in flux. Hence the flares seen in the lightcurves are not merely due to a change in the absorption column density. The electron temperature shows an increase from the second to the last flux level, even though the trend seen is within a few keV. \\
%The N$_{H}$ and $\xi$ of the ionised absorber from method 1 ranges from 30 $\times$ 10$^{22}$ to 80 $\times$ 10$^{22}$ cm$^{-2}$ and 10 to 100 erg~s$^{-1}$~cm, thus indicating that the absorption is approaching a Compton thick state. 

Since the inclination angle is not constrained from spectroscopy alone, the change in the geometry of the reflector with respect to the flux will be an important perspective for the upcoming Imaging X-ray Polarimetry Explorer (IXPE) observations \citep{Weisskopf_2016SPIE.9905E..17W}. With the launch of IXPE late in 2021, the polarisation measurements in the low flux state of this source would be possible, if the source remains in this state until then. X-ray polarimetry will not only assist to substantiate the spectral parameters but it will also provide fundamental information regarding the geometry of the reflectors. A drop in the value of polarisation degree at Fe K$\alpha$ and Fe K$\beta$ lines will indicate if the reflecting matter is neutral and hence the level of ionisation can be probed with polarimetry. From this work we see that the opening angle of the reflector can also be constrained from the polarisation degree. Moreover smaller values of R can give a higher polarisation, and hence the distance to the reflector can also be probed to an extent with polarimetry. However, in OS the continuum is highly absorbed in 2-8 keV range and hence the properties of the inner disk and corona cannot be probed with polarimetry.\\
We simulated the expected IXPE observations using the simulation tool \textit{ixpeobssim} \citep{Rollins_2019NIMPA.936..224P} for the lowest and highest flux levels, and for two cases of the half opening angle (30$^{\circ}$ and 75$^{\circ}$). For the simulation we used the absorbed spectral model derived from the \textit{NuSTAR} data in the 1-10 keV energy range. For the polarisation degree we used the results from section 4.4 for R = 0.1 and N$_{H,refl}$ = $10^{24}$ cm$^{-2}$. For the polarisation angle, we assumed 90$^{\circ}$ and 0$^{\circ}$ for the wind and disk dominating energy ranges. The simulation results from all the three Detector Units (DUs) are added in order to calculate the polarisation degree and angle in four energy bins for the two flux levels and the two opening angles. The results are shown in Fig. \ref{ixpeobssim_fig}. This shows that with a 250 ks exposure of IXPE, we will be able to constrain the opening angle of the absorber and therefore providing the missing piece to the puzzle of GRS 1915+105.

\begin{acknowledgements}
      We thank the anonymous referee for useful comments which helped us to significantly improve the paper. AR thanks Sudip Chakraborty, Riccardo Ferrazzoli and Keigo Fukumura for the useful comments and discussions. This research have made use of the archival data from the NASA's High Energy Astrophysics Science Archive Research Center (HEASARC; \url{https://heasarc.gsfc.nasa.gov/}). We have also used the HEASoft software developed by HEASARC. The Italian contribution to the IXPE mission is supported by the  Italian Space Agency (ASI) through the contract ASI-OHBI-2017-12-I.0, the agreements ASI-INAF-2017-12-H0 and ASI-INFN-2017.13-H0, and its  Space Science Data Center (SSDC), and by the Istituto Nazionale di  Astrofisica (INAF) and the Istituto Nazionale di Fisica Nucleare  (INFN) in Italy.
\end{acknowledgements}

%\end{acknowledgements}
\bibliographystyle{aa}
\bibliography{grs1915absorbedstate}

\end{document}